# SELF-ORGANIZED PULSES OF DYNAMIC SYSTEMS WITH THREE-DIMENSIONAL PHASE SPACE

Z.Zh. Zhanabayev, N.Y. Almasbekov, Y.Zh. Baibolatov, A.T. Yeldesbay

al'Farabi Kazakh national university, Almaty

The possibility of production of pulses with scale-invariant properties at presence of fluctuations of parameters of self-oscillatory system with three-dimensional phase space is shown. The system of equations of inertial nonlinearity oscillator is numerically investigated. The formulas of quantitative characteristics of the form and self-similar values of informational entropy of complex pulses allowing to extract self-organized signals are obtained.

## 1. Introduction

The search of new perspective informational technologies of radio engineering and electronics is closely connected with investigation of dynamic chaos [1]. One of the main properties of real systems with chaotic behavior is the possibility of self-organization in them – appearance of order from chaos. Self-organization is possible at presence of nonlinearity, nonequilibrium and closure failure and has fractal, scale-invariant regularities.

Complex generators stochastic oscillations amplitude-time curve $x(t)$ may also have a fractal structure. It is necessary to take into account that fractality of a random process $x(t)$ has a self-affine (similarity factors with respect to $x$ and $t$ are different), but not a self-similar (similarity factors with respect to $x$ and $t$ are identical) character. Determination of dimensionalities of self-affine fractals in a general view is a non-solved problem.

Fractal dimensionality of the curve $x(t)$ is its metric characteristic. Informational entropy of distribution of points along the curve $x(t)$ is a more general both metric and topological characteristic. At self-organization entropy decreases as compared to its value in an equilibrium state, that's why a normalized entropy may serve as a criterion of degree of self-organization.

For practical applications it is necessary to clear out the following questions: in what way to create and theoretically describe the operation of self-organized signals oscillator, what are the quantitative informational and entropic criteria of the degree of diverse form pulse self-organization (final length signals)? The search of answers to these questions is the aim of the present work.

## 2. Informational criteria of degree of self-organization

According to works [2-4] we'll use the following definitions of information and informational entropy.
Let
$$X = \{x_1, \ldots, x_i, \ldots, x_N\}, \quad Y = \{y_1, \ldots, y_j, \ldots, y_N\} \qquad (1)$$

- be sets of variables characterizing states of corresponding systems designated by the same letters $X, Y$. If $P(x_i / y_i)$ - is the probability of that at state of $x_i$ of system $X$ the system $Y$ transfers to system $y_j$, then the information received by $Y$ is equal to

$$J(x_i / y_i) = -\ln P(x_i / y_i). \tag{2}$$

Conditional informational entropy $S(x/y)$ is determined by averaging of (2) with respect to all states $X, Y$:

$$S(X/Y) = <J(X/Y)> = -\sum_{1 \leq i \leq N} \sum_{1 \leq j \leq N} P(x_i, y_j) \ln(P(x_i / y_j)). \tag{3}$$

Information delivered from $X$ to $Y$, is equal to difference between the starting uncertainty $S(X)$ and final uncertainty $S(X/Y)$:

$$J(X \to Y) = S(X) - S(X/Y) = \sum_{1 \leq i \leq N} \sum_{1 \leq j \leq N} P(x_i, y_j) \ln\left(\frac{P(x_i, y_j)}{P(x_i)P(y_j)}\right). \tag{4}$$

For continuous variables $x$, $y$ formulas (2), (4) are of the form

$$J(x/y) = -\ln P(x/y), \tag{5}$$

$$J(X \to Y) = \iint f(x, y) \ln\left(\frac{f(x, y)}{f(x)f(y)}\right) dx dy, \tag{6}$$

where $f(x), f(y), f(x, y)$ – various densities of probability distribution. At statistical independence $X$, $Y$ we have $f(x, y) = f(x)f(y)$, consequently $J(X \to Y) = 0$.

Formulas (5) and (6) are assumed as various definitions of information. From formula (3) follows that the information determined by formula (5), may be named as random entropy [2], and entropy $S(X/Y)$ – average value of information $J(X/Y)$.

In each of these two definitions (formulas (5), (6)) have the following general properties of information: information is a positive value ($J>0$) and is determined at presence of some condition, asymmetry ($X \neq Y$). Taking this into account we may assume information $J$ as a determining variable having two properties above independent of the method of its determination. We'll specify its additional properties – existence of values determining random (local) and average (global) entropies as interval boundaries of symmetry breaking – self-organization.

Thus, it is possible to speak about probability of information realization:

$$P(J) = e^{-J}, \quad P(J) = \int_0^\infty f(J) dJ, \quad \int_0^\infty f(J) dJ = 1. \tag{7}$$

From this it follows that probability function of information realization $P(J)$ coincides with function of probability density distribution $f(J)$. Just the information may serve as a general and complete characteristic of all hierarchic levels of a complex system: part may contain all information about the whole. Informational entropy is determined as an average value of information:

$$S(J) = \int_J^\infty J f(J) dJ = (J + 1)e^{-J}. \tag{8}$$

For $0 \leq J \leq \infty$ we have $1 \geq P(J) \geq 0$ and $1 \geq S \geq 0$.

Self-similarity of self-organizing system presupposes correspondence of some characteristic function $g(x)$ with functional equation [5]

$$g(x) = \alpha g(g(x/\alpha)), \tag{9}$$

where $\alpha$ - scale factor. Any continuous function in its fixed point satisfies equation (9). Assuming probability function $P(J)$ and informational entropy $S(J)$ as characteristic functions, we find their fixed points:

$$P(J_1) = J_1, \quad e^{-J_1} = J_1, \quad J_1 = 0{,}567, \tag{10}$$

$$S(J_2) = J_2, \quad (J_2 + 1)e^{-J_2} = J_2, \quad J_2 = 0{,}806. \tag{11}$$

These fixed points are uniquely stable as they are also limits of continuous mappings reached at any initial values of information $J_0$:

$$J_{i+1} = P(J_i), \quad \lim_{i \to \infty} \exp(-\exp(\ldots - \exp(-J_0))\ldots)) = J_1; \tag{12}$$

$$J_{i+1} = S(J_i), \quad \lim_{i \to \infty} \exp(-\exp(\ldots - \exp(\ln(J_0 + 1) - J_0))\ldots)) = J_2. \tag{13}$$

Numbers $J_1, J_2$ were determined as well as the possibilities of their use were specified for description of universal regularities of natural phenomena in works [6,7].

Treatment of sense of numbers $J_1, J_2$ may be different, the most universal from them – expansion of field of application of Fibonachee number ("gold section" of dynamic measure of the system). Number $J_1$ corresponds to informational (local) description, number $J_2$ – to entropic (averaged) description of a complex system. We may be convinced in that taking into account that at $J \ll 1$ from (11) follows (10), and taking into account only the first expansion term of the exponent for $J < 1$ from (11) we have the equation for Fibonachee number $J_3$:

$$J_3^2 + J_3 - 1 = 0, \quad J_3 = 0{,}618. \tag{14}$$

## 3. Intermittent signal generator

Stochastic oscillations with corresponding strange attractor in three-dimensional phase space may be produced by modified oscillator with inertial nonlinearity [8]. The oscillator consists of a selective element, amplifier in feedback circuit, amplifier in nonlinear transducer circuit. Output voltages at specified sections we'll designate via $x$, $y$, $z$, then the system of equations of the oscillator is of the form:

$$\dot{x} = (m - z)x + y - d x^3, \quad \dot{y} = -x, \quad \dot{z} = g(\Phi(x) - z), \tag{15}$$

$$\Phi(x) = \phi(x) \cdot x^2, \quad \phi(x) = 0, \; x \leq 0; \; \phi(x) = 1, \; x > 0,$$

where $m$, $g$, $d$ – are correspondingly parameters of excitation, inertia and transconductance of feedback amplifier. In case of high inertia ($g \to 0$) system (15) independent of kind of function $\Phi(x)$ coincides by form of notation with Van der Pol equations. Nonlinearity effect is in proportion to parameter $d$.

In another limiting case of zero lag of oscillator ($g \to \infty$) system (15) is reduced to equation

$$\ddot{x} - (m - \Phi(x) - 3d x^2)\dot{x} + x = 0 \tag{16}$$

and analogy with Van der Pole equation is reached on condition $\Phi(x) = x^2$. In low inertia mode of oscillator operation feedback nonlinearity may be described via behavior of nonlinear transducer and parameter $d$ may be considered constant or equal to zero.

We are interested in pulses with internal structure which are generated at $g \geq 1$. That's why, for simplification of system (15) parameter $d$ may be not taken into account. But on the strength of that now time constant of nonlinear transducer is less than the period of selective element (oscillatory circuit) it is necessary to take into account fluctuation correlations of its parameters. With this aim in view we'll reconsider derivation of system (15).

According to [8] we write the equation for current in oscillator circuit in the form

$$\frac{dI}{dt} + \frac{R}{L}I + (LC)^{-1}\int\left(I - MG\frac{dI}{dt}\right)dt = 0, \qquad (17)$$

where $L$ – inductance, $R$ – resistance, $C$ – capacity of circuit, $M$ – mutual inductance, $G$ – transconductance of amplifier in feedback circuit, $t$ – time. Let's assume

$$y = -\int\frac{I}{\omega_0}dt,\ G = G_0 - bz,\ \dot{z} = -\gamma z + \varphi(I), \qquad (18)$$

where $z$ – voltage at inertial transducer output, $G_0, \gamma, b$ - parameters, $\varphi(I)$ - nonlinear function describing nonlinear transducer operation.

Let's introduce designations

$$\omega_0^2 = 1/L_0 C_0,\ x = I/\omega_0,\ \tau = \omega_0 t,\ m = \frac{M}{LC}\frac{G_0}{\omega_0} - \frac{R}{L\omega_0},\ g = \frac{\gamma}{\omega_0} \qquad (19)$$

and condition

$$b = \omega_0 LC/M,$$

where $L_0, C_0$ – static values $L, C$. Parameters $L, C, M, R$ are assumed as fluctuating in time. For example,

$$L(t) = L_0(1 + \psi(t, \alpha)),$$

where $\psi(t, \alpha)$ – fluctuation function with random initial phase $\alpha$. At averaging in phase $\alpha$ $\langle\psi(t,\alpha)\rangle_\alpha = 0$, but correlation of parameters shall differ from products of their static values. We'll take into account only more slow, the most simple pair correlations. Determined by formula (19) correlations $\left\langle M, \frac{1}{LC} \right\rangle_\alpha$, $\left\langle R, \frac{1}{L} \right\rangle_\alpha$, $\left\langle LC, \frac{1}{M} \right\rangle_\alpha$ are more complicated as compared to correlation

$$\langle LC \rangle_\alpha = L_0 C_0 K(m, g, \tau), \qquad (20)$$

which is contained in formula (17). Correlation function form $K(m,g,\tau)$ is determined by concrete model of a random process. The above means

$$\left\langle \frac{M}{LC} \right\rangle_\alpha = \frac{M_0}{L_0 C_0},\ \left\langle \frac{R}{L} \right\rangle_\alpha = \frac{R_0}{L_0},\ \left\langle \frac{LC}{M} \right\rangle_\alpha = \frac{L_0 C_0}{M_0},\ \langle LC \rangle_\alpha = L_0 C_0 K(m, g, \tau). \qquad (21)$$

Assuming $\varphi(J) = g\phi(x)x^2$, from (17) and (21) we obtain for derivatives with respect to $\tau$:

$$\begin{cases} \dot{x} = (m-z)x + y/K(m,g,\tau), \\ \dot{y} = -x, \\ \dot{z} = g(\phi(x)x^2 - z), \end{cases} \qquad \phi(x) = \begin{cases} 1, x > 0, \\ 0, x \leq 0. \end{cases} \qquad (22)$$

By sense $K(m,g,\tau) \neq 0$.

Let's determine correlation function of fluctuations in the form of a harmonic signal with random uniformly distributed initial phase and random amplitude:

$$\psi(t, \alpha) = A\cos(\omega t + \alpha). \qquad (23)$$

Average value with respect to $\alpha$ of product $L(t_1)C(t_2)$ is of the form

$$\langle L(t_1)C(t_2)\rangle_\alpha = L_0 C_0\left(1 + \frac{A^2}{2}\cos(\omega(t_2 - t_1))\right). \qquad (24)$$

Assuming random process as stationary $(t_2 - t_1 = t)$, we'll select correlation frequency from condition of self-consistency of a selective element and nonlinear transducer:

$$\frac{\omega}{\omega_0} = \frac{\omega_0}{\gamma}, \quad \omega = \frac{\omega_0}{g}, \quad \omega t = \frac{\tau}{g}. \tag{25}$$

Amplitude of fluctuations $A$ may be assessed from equality of fluctuations of average energy (dispersion) $\frac{A^2}{2}$ and specific in frequency of energy supplied to oscillatory circuit $m/g$. From this it follows also that there is no necessity to search for average value of $\langle A_1(t_1)A_2(t_2)\rangle$, as this correlator is complicated, quick-damped. After this the required correlation function is of the form

$$K(m,g,\tau) = 1 + \frac{m}{g}\cos\frac{\tau}{g}. \tag{26}$$

By sense of problems considered $K(m,g,\tau) \neq 0$, consequently, the condition $m < g$ should be satisfied. This inequality is also the condition of dissipativity of a self-oscillatory system without an inertial transducer. Actually, writing system (22) in a vector form

$$\dot{\vec{x}} = \vec{F}(\vec{x}, m, g), \tag{27}$$

we have $div\vec{F} = \left\{\frac{\partial \dot{x}}{\partial x} + \frac{\partial \dot{y}}{\partial y} + \frac{\partial \dot{z}}{\partial z}\right\} < 0$ at $m < g, z = 0$.

From linear analysis of stability of system (22) follow conditions of excitation

$$\sqrt{2} < m < g, \tag{28}$$

where estimate $K_{max}(m,g,\tau) = 1 + \frac{m}{g}$ is taken.

## 4. Pulse shape factors

The informational entropy is a metric and topological characteristic. In order to quantitatively describe its regularities it is necessary to use another, for example, purely metric characteristic. With this aim in view we'll determine more general metric characteristic unambiguously describing the difference of some pulses shapes. We'll make use of the fact that the existence of metric characteristics (length, area, volume) follows from fulfilling Cauchy-Bunyakowsky inequality:

$$\left(\frac{1}{T}\int_0^T x^2(t)dt\right)^{1/2} \geq k_1\left(\frac{1}{T}\int_0^T |x(t)|dt\right) \quad \text{или} \quad \langle x^2 \rangle^{1/2} \geq k_1\langle |x|\rangle, \tag{29}$$

where $t$ and $T$ may have the sense of a current and characteristic time. Inequality is satisfied at

$$k_1 = \frac{\langle x^2(t)\rangle^{1/2}}{\langle |x(t)|\rangle} = const. \tag{30}$$

Quantity $k_1$ is used in radiophysics and is named pulse signal form factor.

Inequality (29) follows from an integral Gelder inequality for any functions $x_1(t)$, $x_2(t)$ written in the form

$$\left(\frac{1}{T}\int_0^T |x_1(t)|^p dt\right)^{1/p} \left(\frac{1}{T}\int_0^T |x_2(t)|^q dt\right)^{1/q} \geq k_{p,q} \frac{1}{T}\int_0^T |x_1(t)x_2(t)| dt, \quad \frac{1}{p}+\frac{1}{q}=1, \quad (31)$$

where $k_{p,q}$ – factor at constant value of which equality in (31) is satisfied. At $x_1(t) \equiv x(t)$, $x_2(t) \equiv 1$ we'll have formula (30). For case $p = q = 2$ in (31) equality is satisfied at

$$k_2 = \frac{\left(\langle x_1^2(t)\rangle\right)^{1/2}\left(\langle x_2^2(t)\rangle\right)^{1/2}}{\langle |x_1(t)x_2(t)|\rangle} = const. \quad (32)$$

Introduction of $k_2$ quantity is necessary for description of self-affine curves. The thing is that the known characteristics, as, for example, metric characteristic distance and statistical characteristic dispersion determined even by two dimensionless variables $x_i, t_i$ do not take into account asymmetry of the curve $x(t)$ with respect to parameter (time) $t$ in contrast to $k_2$ quantity.

## 5. Results of numerical analysis

System of equations (22), (26) was numerically investigated by Runge-Kutta method. In comparatively narrow intervals of parameters values $m \in [1.42 \div 1.65]$, $g \in [1.45 \div 5.00]$ various kinds of oscillations are realized: quasi-periodic-, noise- and intermittent oscillations. The informational entropy of pulses and their set (signal) were determined by the following formula:

$$S(\delta) = -\sum_{i=1}^{N} P_i(\delta) \ln P_i(\delta), \quad P_i(\delta) = \lim_{N \to \infty} \frac{N(\delta_i)}{N}, \quad (33)$$

where $\delta$ – size of cells of maximum amplitudes partition $x(t)$, $y(t)$, $z(t)$, $P_i$ – probabilities of revealing of a number of points $N(\delta_i)$ in a cell with number $i$. In main calculations it is assumed $\delta = 10^{-3}$.

In figs. 1 and 2 are presented an example of realizations, corresponding to it phase portrait and extracted pulses. As pulses there were chosen signal fragments between zero amplitude values as it is shown in fig.1, where $\nu$ – number of zeroes $x(t)$ within a pulse. Factors of the form $k_2(x)$, $k_2(y)$, $k_2(z)$, determined in accordance with formula (32) in the form

$$k_2(x,t) = \frac{\left(\langle x^2(t)\rangle\right)^{1/2}\left(\langle t^2\rangle\right)^{1/2}}{\langle |x(t)\cdot t|\rangle} \quad (34)$$

unambiguously differentiate kinds of oscillations for identical values $\nu$. At $\nu = 0$ for rectangular pulse $k_2 = 1$, for triangular pulse – $k_2 = k_{2m} = 1,33$. Pulses of more complicated form have the values $k_2 > k_{2m}$. At $\nu = 1$ pulse of intermittent (with alternation of bursts and small-scale oscillations) signal $x(t)$ has the greatest value $k_2(x) = 5,683$, and $k_2(y) = 2,413$, $k_2(z) = 1,958$ $k_2(z)=1,958$. For signals with $\nu \gg 1$ value $k_2$ is stabilized about a half of its maximum.

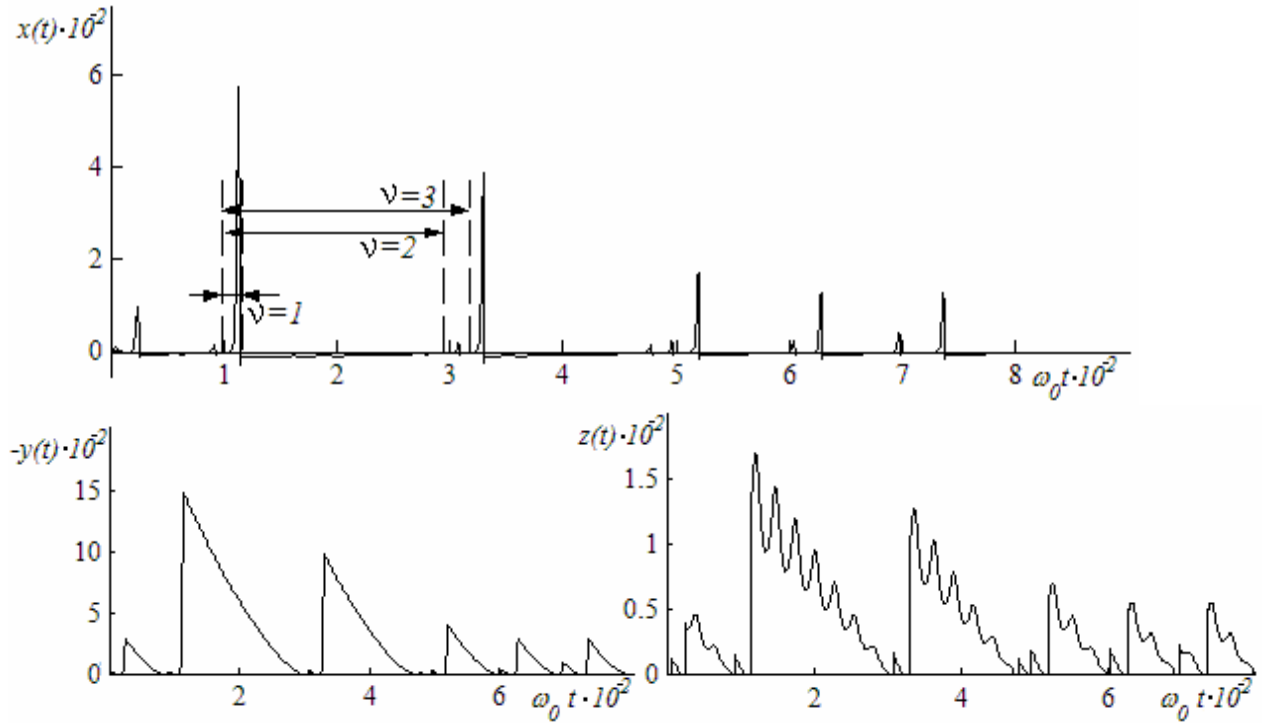

Fig. 1. Realizations *x(t)*, *y(t)*, *z(t)* at *m*=1.6, *g*=4.3

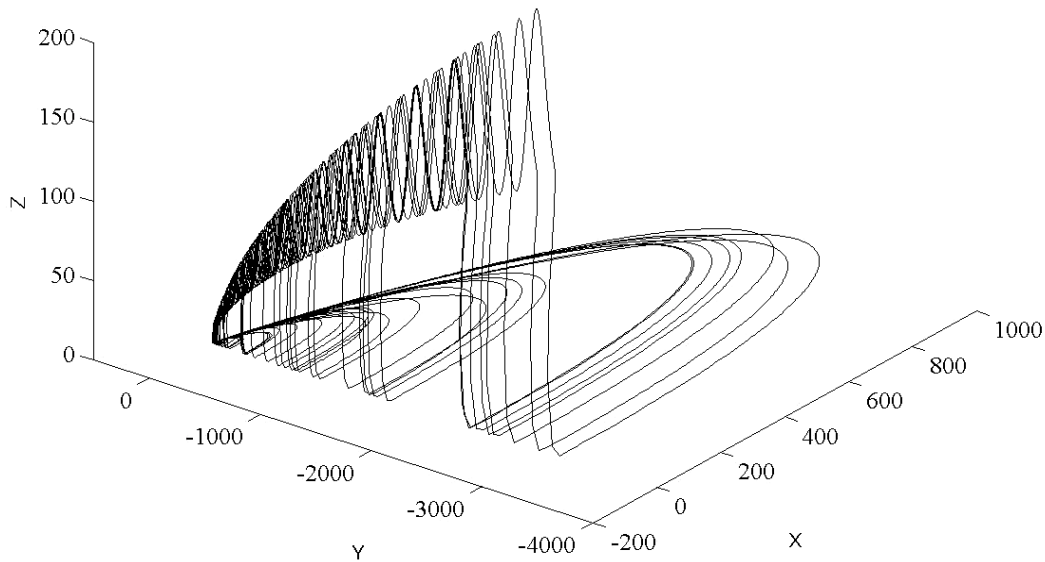

Fig. 2. Phase portrait of oscillations at *m*=1.6, *g*=4.3.

In fig. 3 the change of a relative value of informational entropy of pulses depending upon form factor is shown. The maximum entropy $S_m = 7,451$ at $\delta = 10^{-3}$ corresponds to isosceles triangular pulse. Only intermittent signals *x(t)*, consisting of positive and negative pulses, belong to self-organization interval $S/S_m \in [J_1, J_2]$, where $J_1, J_2$ are determined by formulas (10), (11). Values $1 \lesssim k_2/k_{2m} \lesssim 2.5$ correspond to various pulses, including intermittent with $\nu = 1, 2, 3$, and values $2.5 \lesssim k_2/k_{2m} \lesssim 8$ — only to intermittent signal with $\nu = 1$.

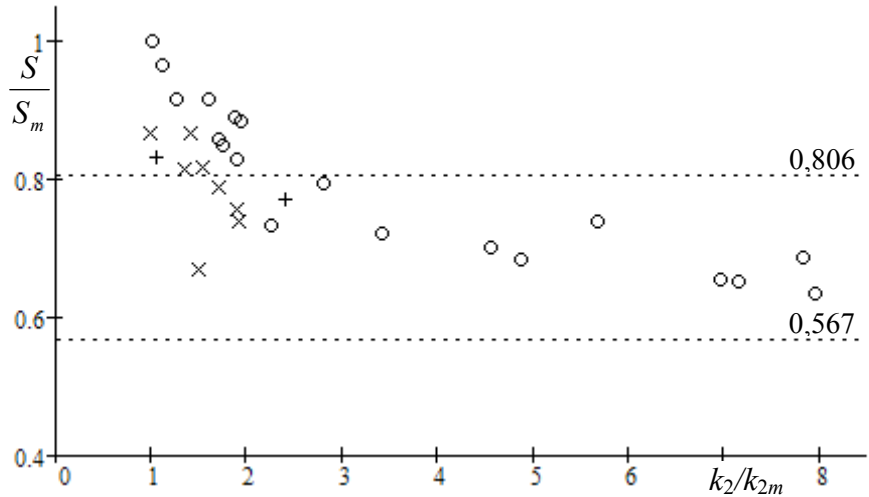

Fig. 3. Dependence of pulse relative entropy on form factor. $\nu$ : × – 0, ○ – 1, + – 2.

Scale invariance of self-organized signals may be directly observed from Fig. 1: pulses $x(t)$ with equal $\nu$ values have a similar form. We'll note that to above specified self-organization criteria ($S/S_m \in [J_1, J_2]$, $k_2/k_{2m} \gtrsim 2{,}5$) correspond forms of electrocardiograms of a healthy man (fig. 4a), and cases of heart disease (fig. 4b) are characterized by values $S, k_2$ [9].

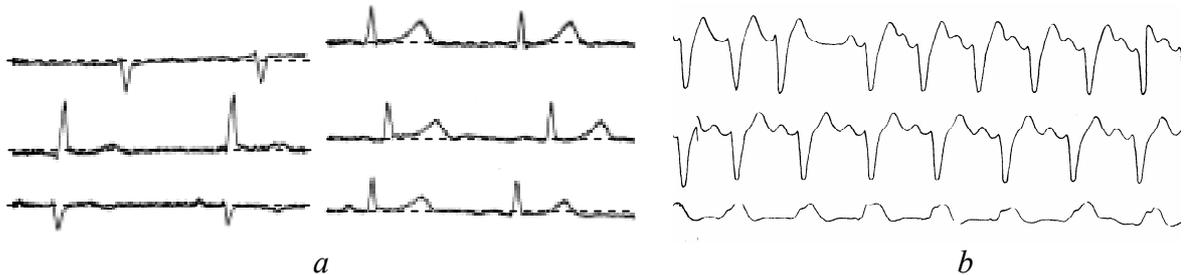

*a*            *b*

Fig. 4. Electrocardiograms of a healthy (*a*) and sick (*b*) heart of a man.

## 6. Conclusion

Account of chaotic motion component in the simplest nonlinear system (with 1.5 degrees of freedom) allows to quantitatively describe the possibilities of producing signals with scale-invariant properties. The informational and entropic criteria of self-organization degree and general metric characteristic of self-affine (with various similarity factors with respect to diverse variables) pulses are the key values specified by us. The results obtained may have wide applications for investigation of complicated systems of diverse nature.